\documentclass[11pt,a4paper]{article}

\usepackage[margin=1in]{geometry}
\usepackage{amsmath,amssymb}
\usepackage{booktabs}
\usepackage{graphicx}
\usepackage{xcolor}
\usepackage{enumitem}
\usepackage{array}
\usepackage{tabularx}
\usepackage{float}
\usepackage{url}
\usepackage{hyperref}
\usepackage[numbers]{natbib}
\usepackage{authblk}
\usepackage[font=small,labelfont=bf]{caption}

\makeatletter
\DeclareRobustCommand{\code}[1]{%
  \begingroup
  \def\_{\textunderscore\allowbreak}%
  \texttt{#1}%
  \endgroup}
\makeatother

\hypersetup{
    colorlinks=true,
    linkcolor=blue!70!black,
    citecolor=blue!70!black,
    urlcolor=blue!70!black
}

\begin{document}

\title{Control Physiology: An Agent-Based Model of FAIR-CAM Dynamics}

\author[1]{Jack Jones}
\author[2]{Laura Voicu\thanks{Corresponding author: laura.voicu@me.com}}
\affil[1]{Enterprise Risk Quantification Institute, USA}
\affil[2]{Enterprise Risk Quantification Institute, Switzerland}

\date{}

\maketitle

\begin{abstract}
Security risk analysis typically treats control effectiveness as a static input, yet controls degrade through configuration drift, depend on monitoring systems that may themselves be degraded, and compete for finite remediation budgets. The FAIR Controls Analytics Model (FAIR-CAM) provides the theoretical framework for these dynamics, decomposing controls into three interacting domains (Loss Event Controls, Variance Management Controls, and Decision Support Controls) but has so far remained theoretical.

We present the first agent-based model to operationalize the core FAIR-CAM dynamics, making control physiology computationally observable, and are releasing the implementation as open source. The simulation implements eight agent types, a multiplicative defense-in-depth susceptibility formula, a three-source variance model, budget-constrained remediation, and a narrative causation engine that produces a complete causal trace for every loss event. In a hospital ransomware scenario ($N=1{,}000$ iterations), three organizational dynamics emerge that static analysis cannot represent. First, emergent operational efficacy diverges from the analytical FAIR-CAM formula by approximately 17\%, driven by correlated extrinsic variance; the divergence grows linearly with extrinsic frequency and vanishes under purely intrinsic drift. Second, a sharp queueing regime transition in the remediation pipeline approximately $2.8\times$ expected loss when budget falls below a scenario-specific threshold (5--10 engineer-hours/month in this parameterization). Third, cascading monitoring failures propagate through the VMC topology: a single degraded VMC silently compounds undetected variance across the controls it manages. These dynamics (analytic divergence under correlated variance, budget-driven regime transitions, and silent monitoring cascades) are structural properties of the FAIR-CAM architecture and should generalize beyond the specific scenario studied.
\end{abstract}


\section{Introduction}

\subsection{The gap between control anatomy and control physiology}

Security risk quantification has historically treated controls as static parameters, describing what a control \emph{is} (its anatomy) without modeling how it \emph{works} over time (its physiology). Whether expressed as point estimates (``the firewall is 85\% effective'') or probability distributions (``efficacy is Beta-PERT distributed between 0.7 and 0.95''), the conventional approach captures anatomy only. In practice, controls degrade through configuration drift, personnel turnover, and evolving threats. They are managed by other controls, operated by people who make imperfect decisions, and constrained by budgets that may be inadequate for the remediation workload. These dynamic properties shape the risk posture that controls collectively provide.

The cybersecurity industry has built increasingly sophisticated models of the technical kill chain (how attacks progress through initial access, lateral movement, and exfiltration). Cyber ranges simulate these attacks at high fidelity. Penetration testing frameworks model attacker capabilities. Threat intelligence feeds track adversary TTPs. None of these tools model the \emph{organizational} dynamics that determine whether controls actually work: decision-making quality, control variance over time, the interplay between policies and operational behavior, and the causal chains from management decisions to loss events.

The FAIR Controls Analytics Model (FAIR-CAM), developed by Jones and published by the FAIR Institute~\citep{Jones2023FAIRCAM} as an extension of the Factor Analysis of Information Risk (FAIR) framework, addresses this gap by introducing control \emph{physiology}: a systematic model of how controls function, degrade, and recover over time. FAIR-CAM decomposes the control ecosystem into three domains: Loss Event Controls (LECs) that directly resist, detect, and respond to threats; Variance Management Controls (VMCs) that manage degradation in other controls through variance prevention, detection, and correction; and Decision Support Controls (DSCs) that influence the quality of personnel decisions about control maintenance and operation.

This three-domain structure creates indirect, time-dependent dependencies that are \emph{organizational}, not technical (Section~\ref{sec:faircam} details each domain). These are the dynamics that static risk analysis cannot capture, and no tool implements the full FAIR-CAM model. This paper presents such an implementation.

\subsection{Why agent-based modeling}

FAIR-CAM's mechanics are dynamic: controls drift according to stochastic change-frequency distributions, personnel make decisions gated by probabilistic alignment models, threats arrive as Poisson processes, and remediation is constrained by finite budgets that reset monthly. The analytical formula for operational efficacy --- $\mathrm{OpEff} = \mathrm{Cov} \times [(\mathrm{Rel} \times \mathrm{IntEff}) + ((1-\mathrm{Rel}) \times \mathrm{VarEff})]$ --- assumes stationarity: that reliability and efficacy are time-invariant averages. This assumption holds when controls are independent, remediation capacity exceeds demand, and variance management is reliable. Under these conditions, the formula is both correct and sufficient. But in real organizations, these conditions frequently fail.

Agent-based modeling (ABM) provides a natural computational framework for FAIR-CAM because each component of the model (each control, each threat, each person) can be represented as an autonomous agent with its own state machine, decision rules, and interactions. In our implementation, reliability is an emergent property: the fraction of time a control spends in its intended state over the simulation run. Operational efficacy emerges similarly as the time-weighted average of intended and variant efficacy. Loss is the cumulative result of all breach events, each resolved through the full contact-breach-detection-response pipeline. This emergence captures temporal dynamics that analytical formulas cannot: variance accumulation during budget shortfalls, and cascading detection failures when VMCs are themselves variant. The simulator also implements a dynamic personnel-behavior model (social contagion, satisficing, event shocks); that component is disabled for all experiments in this paper (see Section~\ref{sec:knownlim}).

The agent-based modeling approach also enables a capability unique to simulation: narrative causation tracing. Each loss event can be traced backward through the causal chain (which controls were variant, why, which VMC failed to detect the variance, which personnel decision allowed it) producing actionable root cause analysis that connects financial loss to specific organizational dynamics.

\subsection{What we model and what we do not}

A key modeling choice in this work: we model organizational dynamics, not network architecture. The simulation represents structural relationships between controls, the people who manage them, the monitoring processes that maintain them, and the budget constraints that limit remediation. Scale is captured in parameters, not agent counts: ``Personal Workstations'' represents a fleet of hundreds of devices whose parameters are calibrated to reflect the population. This enables Monte Carlo batch analysis that would be prohibitively expensive at per-endpoint granularity. The abstraction philosophy is detailed in Section~\ref{sec:abstraction}.

\subsection{Contribution}

This paper makes four contributions:
\begin{enumerate}[nosep]
    \item We present the first computational implementation of the core FAIR-CAM dynamics, making the organizational processes described by the framework computationally observable.
    \item We observe in simulation that these organizational dynamics produce non-linear risk behavior: a remediation budget threshold exists below which variance accumulation exceeds remediation capacity, approximately tripling expected annual loss in our hospital scenario.
    \item We describe a narrative causation engine that traces every loss event to its root organizational causes (specific variance events, the monitoring failures that let them persist, and the budget/queue state that gated their remediation), enabling automated root cause analysis within the simulation.
    \item We validate the model against FAIR-CAM's analytical formulas, showing convergence under stationary conditions and characterizing the divergence under realistic organizational constraints (budget limitations, cascading VMC failures, correlated extrinsic variance).
\end{enumerate}

\section{Background}

\subsection{The FAIR taxonomy}

The Factor Analysis of Information Risk (FAIR) provides a structured decomposition of cyber risk into quantifiable components \citep{Freund2015}. At its core, risk is expressed as the product of Loss Event Frequency (LEF) and Loss Magnitude (LM). LEF decomposes through a multi-stage attack pipeline rather than a single susceptibility parameter: LEF decomposes through three sequential Prevention gates --- avoidance, deterrence, and resistance --- consistent with the FAIR-CAM Loss Event Control taxonomy. A threat first generates a \emph{Contact} with an asset; the contact may be blocked by an avoidance control, which prevents contact from occurring at all. If the contact is not avoided, it becomes a candidate \emph{Threat Event}; at this point deterrence controls may drive the threat's probability of action toward zero so that, although the threat is in position, it elects not to act. Deterrence presupposes a threat that exercises choice (e.g.\ insider actors). If the threat does act, its capability is compared against the asset's combined resistance strength, and a \emph{Loss Event} occurs only when threat capability exceeds resistance. Susceptibility in the FAIR taxonomy is therefore the conditional probability that a threat event --- a contact that has survived the avoidance gate and, where applicable, the deterrence gate --- results in a loss event by overcoming combined resistance, not simply the probability that any contact does so.

Loss Magnitude decomposes into Primary Loss (the direct consequences to the primary stakeholder: productivity loss during the incident, response costs for forensics and recovery, and replacement costs) and Secondary Loss (the consequences of secondary stakeholder reactions: fines and judgments, reputation damage, competitive-advantage loss, and secondary response costs such as customer notification and credit monitoring), gated by the Secondary Loss Event Frequency (SLEF), the probability that a primary loss triggers secondary consequences.

The simulator implements all three pre-loss gates as separate stages with independent metrics (avoided contacts, deterred events, resisted events).

FAIR's strength is its decomposition: each component can be estimated independently via calibrated expert judgment, empirical data, or both. This enables quantitative risk analysis that produces dollar-denominated loss distributions rather than ordinal ratings. However, FAIR's original decomposition treats control effectiveness as a static input in the form of a distribution for Susceptibility without modeling the dynamic processes that determine how effective controls actually are over time. Moreover, beyond Resistance Strength, FAIR provides no explicit place to account for the many other types of controls --- detection, response, variance management, decision support --- that shape real-world risk outcomes. FAIR-CAM extends FAIR by providing this dynamic model.

\subsection{FAIR-CAM: Controls Analytics Model}
\label{sec:faircam}

FAIR-CAM organizes controls into three domains that operate at different levels of indirection relative to threat events.

\paragraph{Loss Event Controls (LECs)} are the controls that directly interact with threats. They operate in three phases: Prevention, comprising avoidance (making the asset invisible to threats), deterrence (discouraging the threat from acting), and resistance (blocking the threat's action); Detection, requiring all three subfunctions of visibility (making threat activity observable), monitoring (actively looking for threat activity), and recognition (correctly interpreting what is observed); and Response, comprising event containment (stopping the ongoing breach), resilience (recovering operational capability), and loss minimization (reducing the financial impact of the event). Each LEC has an intended efficacy, a variant-state efficacy, and a control type that determines its role in the contact pipeline.

\paragraph{Variance Management Controls (VMCs)} do not prevent breaches directly. Instead, they manage the reliability of other controls by preventing variance, detecting degradation, and implementing corrections. VMC functions include reducing change frequency, reducing variance probability, threat capability monitoring, controls monitoring, treatment selection, and implementation of corrections. A VMC that is itself in a variant state cannot reliably detect variance in the controls it manages, creating the potential for cascading detection failures.

\paragraph{Decision Support Controls (DSCs)} influence the quality of personnel decisions about control operation and maintenance. The DSC model evaluates five dimensions: expectation alignment (do formal expectations exist and have they been communicated?), awareness (does the person know the policy?), capability (can they follow it?), situational awareness (does the context support compliance?), and incentive alignment (are they motivated to comply?). Each dimension is evaluated as a Bernoulli trial; the combination determines the probability of a misaligned decision via a conditional probability table. When personnel are misaligned, the controls they operate provide only partial protection proportional to the dimensions that passed.

The key analytical formulas are:
\begin{align}
    \mathrm{OpEff} &= \mathrm{Cov} \times \bigl[(\mathrm{Rel} \times \mathrm{IntEff}) + ((1-\mathrm{Rel}) \times \mathrm{VarEff})\bigr] \label{eq:opeff} \\
    S &= \prod_{i} (1 - \mathrm{OpEff}_i) \quad \text{(defense-in-depth)} \label{eq:suscept} \\
    \mathrm{Detection} &= V \wedge M \wedge R \quad \text{(AND-gate)} \label{eq:detect}
\end{align}
where Coverage (Cov) represents the fraction of assets protected, Reliability (Rel) the fraction of time in intended state, IntEff the intended efficacy, and VarEff the variant-state efficacy.

\subsection{Related computational approaches}

\paragraph{Cybersecurity ABMs.} Agent-based models have been applied to cybersecurity in several contexts, spanning three broad dimensions.

On the \emph{technical} dimension, \citet{Wagner2015} developed an ABM at MIT Lincoln Laboratory for assessing network security risk due to unauthorized hardware, using agents to model asset discovery and vulnerability exploitation. \citet{Thompson2018} applied ABM to military tactical networks where agents represent network nodes and adversarial actors. Holm et al.\ developed CySeMoL and its probabilistic extension P\textsuperscript{2}CySeMoL \citep{Holm2014}, a Bayesian attack graph framework that models attack success probabilities conditioned on deployed countermeasures. Kotenko and Chechulin \citep{Kotenko2013} built attack modeling and security evaluation frameworks using multi-agent simulation for network-level threat analysis.

On the \emph{defender-decision} dimension, Dutt et al.\ \citep{Dutt2013} applied Instance-Based Learning Theory to model detection of cyber attacks, producing an agent-based model of security analyst behavior under uncertainty, while Ben-Asher and Gonzalez \citep{BenAsher2015} studied how cyber security knowledge affects attack detection within the same cognitive framework. \citet{Rausch2018} proposed GAMES (General Agent Model for the Evaluation of Security), a formalism for composable security agent models providing a modular framework where security components are represented as interacting agents. \citet{Cernohorsky2018} demonstrated ABM viability for cybersecurity scenario analysis more broadly.

None of these approaches model the \emph{organizational control maintenance} dimension: monitoring, variance management, and remediation processes that determine whether controls sustain their intended effectiveness over time.

\paragraph{Risk quantification and economics.} Game-theoretic approaches model optimal control portfolios under budget constraints. \citet{Fielder2016} and \citet{Panaousis2014} produce investment recommendations that optimize \emph{which} controls to deploy but treat each control's effectiveness as a static parameter. \citet{Gordon2002} established the foundational economic model for security investment. Anderson et al.\ \citep{Anderson2013} provided comprehensive measurement of cybersecurity costs, establishing empirical baselines for loss magnitude calibration. These approaches optimize which controls to deploy without modeling how they degrade and recover over time.

\paragraph{Safety science and organizational failure.} Our organizational framing draws on a mature literature in safety science. Reason's Swiss cheese model \citep{Reason1990} introduced the concept of latent conditions --- dormant organizational failures that combine with active errors to produce accidents --- which directly parallels the VMC cascade mechanism where dormant monitoring failures enable active control degradation. Perrow's normal accidents theory \citep{Perrow1984} describes how cascading failures in tightly coupled systems produce outcomes that no single component failure could explain, a property our monitoring topology exhibits. Rasmussen's dynamic safety model \citep{Rasmussen1997} and Dekker's drift-into-failure framework \citep{Dekker2011} describe how systems migrate toward failure boundaries through incremental, locally rational decisions --- the organizational dynamic our personnel behavior model is designed to capture. Leveson's STAMP framework \citep{Leveson2012} takes a systems-theoretic view where safety constraints are enforced by control structures that can degrade; FAIR-CAM's three-domain control taxonomy (LEC/VMC/DSC) can be read as a domain-specific instantiation of this principle. Hollnagel's barrier analysis \citep{Hollnagel2004} provides a taxonomy of defense degradation modes (absent, bypassed, ineffective) that maps to FAIR-CAM's variance ontology.

\paragraph{Positioning.} None of these approaches provide an integrated computational model of the organizational processes that sustain control effectiveness: monitoring quality, decision alignment, remediation capacity, and the cascading interdependencies between them. A recent meta-review \citep{Woods2024} confirms that the evidence base for control effectiveness is thin and largely cross-sectional; longitudinal or simulation-based evidence is effectively absent. Our work fills this gap by operationalizing FAIR-CAM's three-domain taxonomy as an agent-based simulation. The contribution is not the individual mechanisms (queueing, cascading failure, and personnel decision modeling all have mature literatures) but their integration into a single, parameterizable, open-source framework aligned with the FAIR-CAM standard.

\section{Model Architecture}

\subsection{Abstraction philosophy}
\label{sec:abstraction}

Our model abstracts away technical detail (individual endpoints, packet flows, exploit mechanics) and invests its fidelity in organizational dynamics (control variance, monitoring reliability, personnel decisions, remediation capacity). This is consistent with how simulation is used across scientific disciplines: epidemiological models do not model every individual human; flight simulators do not model every molecule of air; climate models do not simulate every raindrop. The fidelity is in the \emph{dynamics} of the model.

We add granularity where it earns its keep. Hospital Floor Workstations are separated from Personal Workstations because they have different exposure profiles and different relationships to business assets. But we do not separately model every individual workstation because they are functionally identical from a risk perspective. Shadow IT is accommodated as an explicit technology-asset class with higher contact frequency and no associated controls.

\subsection{Agent types}

The simulation implements eight agent types, summarized in Table~\ref{tab:agents}.

\begin{table}[htbp]
\centering
\caption{Agent types, roles, key state variables, and interactions.}
\label{tab:agents}
\small
\begin{tabularx}{\textwidth}{l l X X}
\toprule
\textbf{Agent} & \textbf{Role} & \textbf{Key State} & \textbf{Interactions} \\
\midrule
ThreatSource & Schedules contacts (Poisson) & Contact frequency, sophistication dist. & Creates ThreatAgents \\
ThreatAgent & Executes attack stages & Sophistication, origin, velocity & Contacts TechAssets \\
TechAsset & Hosts business assets & Network layer, compromised flag & Protected by LECs \\
BusinessAsset & Carries value & Asset type, value class & Hosted by TechAssets \\
LEC & Resists, detects, terminates & Efficacy (int/var), state, control type & Monitored by VMCs \\
VMC & Monitors controls & Sweep interval, detection method & Monitors LECs/ VMCs/ DSCs \\
DSC & Influences decisions & Efficacy per dimension & Influences Personnel \\
Personnel & Manages controls & Propensity, CVF profile, DSC attrs & Operates LECs/ VMCs/ DSCs \\
\bottomrule
\end{tabularx}
\end{table}

Each technology asset agent represents a \emph{class} of infrastructure, not an individual device. ``Personal Workstations'' in the hospital scenario represents a fleet of hundreds of devices; its parameters (contact frequency, lateral movement connectivity, and the controls that protect it) are calibrated to reflect the population. This is how scale enters the model: through parameters, not agent counts. If an organization distinguishes DMZ servers from internal servers because they have different control configurations and exposure profiles, we add a second technology asset agent. The test is whether the distinction changes how risk flows through the system.

\subsection{Simulation loop}

The simulation operates in discrete hourly ticks. Each tick represents an hour and executes a nine-step pipeline that processes all agent interactions in a fixed order. The ordering reflects the FAIR-CAM causal sequence: variance sources fire first (extrinsic and intrinsic), then VMC detection sweeps can discover newly variant controls, then pending losses are realized, then remediation completes any controls whose repair duration has elapsed, and finally threat contacts are processed against the current control state. Remediation precedes contacts so that a control completing repair in tick $t$ contributes its intended efficacy to any contact resolved in the same tick --- a modeling choice that favours the defender at the one-hour resolution. The experiments reported in this paper use three horizons depending on the question asked: 8,760 ticks (1 year) for Monte-Carlo batch and sensitivity analyses, 26,280 ticks (3 years) for backlog-dynamics traces, and 43,800 ticks (5 years) for monitoring-cascade traces. Each figure and table caption states the horizon used.

\begin{enumerate}[nosep]
    \item Personnel behavior update (propensity, social influence, event shocks)
    \item OpEx accumulation
    \item Extrinsic variance (threat landscape Poisson events)
    \item Intrinsic variance (drift timers + monthly personnel checks)
    \item VMC detection sweeps
    \item Realize pending losses (detection $\to$ termination $\to$ recovery $\to$ magnitude)
    \item Remediation queue processing (budget-gated)
    \item Threat contact processing (targeting $\to$ avoidance $\to$ deterrence $\to$ resistance $\to$ breach/\allowbreak resist)
    \item Metrics collection
\end{enumerate}

\subsection{Organizational structure}

The organizational structure is represented as a directed multigraph where nodes are agents and edges carry typed relationships. Nine relationship types define the structure:

\smallskip
\noindent\begin{tabular}{@{}ll@{}}
\texttt{LEC\_PROTECTS\_ASSET} & Which assets a control defends \\
\texttt{VMC\_MONITORS} & Which controls a VMC monitors for variance \\
\texttt{DSC\_INFLUENCES} & Which personnel a DSC affects \\
\texttt{TECH\_CONNECTS\_TECH} & Network adjacency (lateral movement) \\
\texttt{BA\_HOSTED\_BY\_TA} & Business asset hosting \\
\texttt{PERSONNEL\_OPERATES} & Control ownership linkage \\
\texttt{VMC\_THREAT\_INTEL} & Threat intelligence feed \\
\texttt{VMC\_SELECTS\_TREATMENT} & Remediation prioritization \\
\texttt{VMC\_IMPLEMENTS} & Remediation implementation gating \\
\end{tabular}
\smallskip

\begin{figure}[htbp]
\centering
\includegraphics[width=0.95\textwidth]{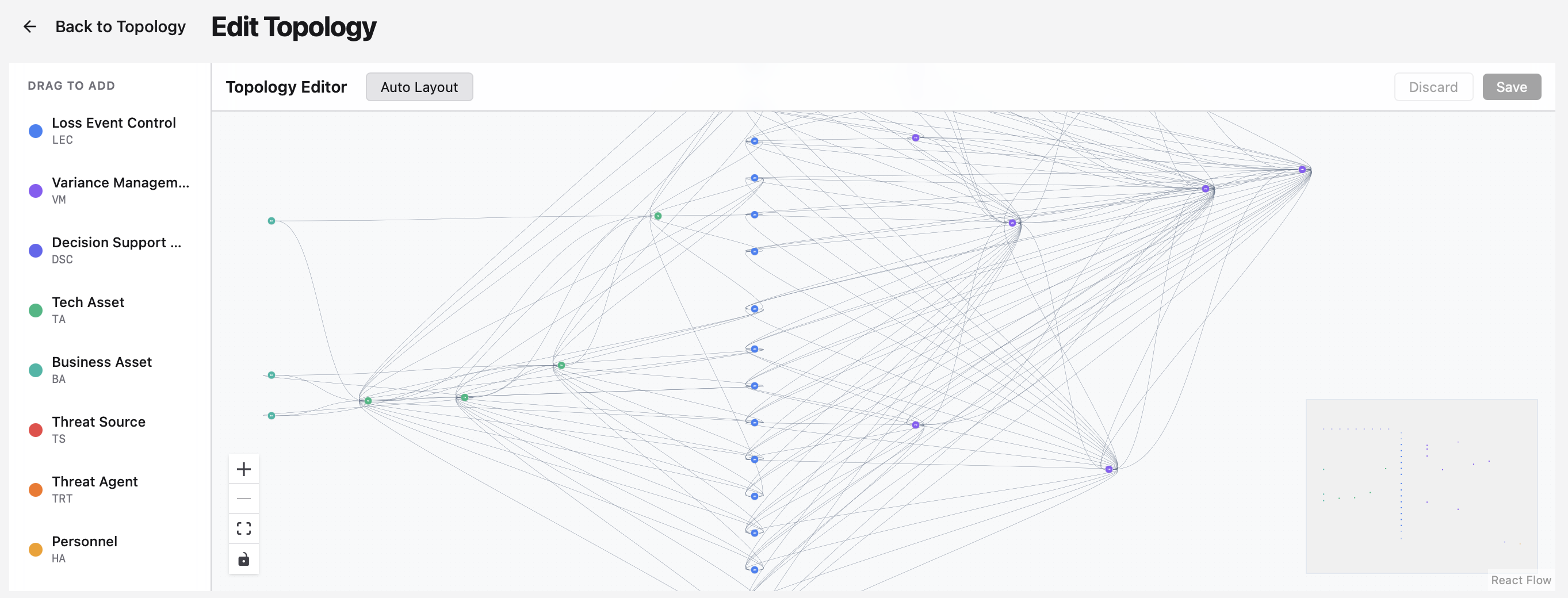}
\caption{Hospital medium-posture organizational structure in the visual editor. Color-coded nodes (blue = LECs, teal = VMCs, purple = DSCs, green = tech/business assets, red = threats, orange = personnel) connected by typed relationship edges. The density of connections reflects the organizational interdependencies that drive the dynamics described in this paper.}
\label{fig:topology}
\end{figure}

The organizational structure serves as a ``digital twin'' and it captures how controls are wired together: which VMCs monitor which LECs, which personnel manage which controls, which business assets are exposed if a particular control degrades. These relationships determine the \emph{organizational} dynamics: cascading VMC failures propagate through monitoring edges, personnel-driven variance propagates through ownership edges when the dynamic personnel-behavior model is enabled, and remediation bottlenecks affect controls in queue-priority order. The same organizational structure can be subjected to different threat actors, frequencies, and sophistication levels. Figure~\ref{fig:topology} shows the hospital medium scenario in the visual editor. All experiments in this paper use two parameterizations of this hospital topology --- a \emph{medium-controls} variant (adequately resourced) and a \emph{weak-controls} variant (deliberately stressed to exercise the variance dynamics) --- described in Section~\ref{sec:scenarios}.

\subsection{Hospital ransomware scenarios}
\label{sec:scenarios}

All experiments in this paper use a hospital ransomware setting with two scenario parameterizations of the organizational structure above, designed to bracket the range from inadequately to adequately resourced controls.

\paragraph{Scenario A: Weak controls.} 14 LECs with low resistance efficacy (five resistance LECs at 20--70\% intended efficacy), no SIEM, no OS hardening, slow VMC detection (monthly to annual sweeps), weak DSCs (30--60\%), one under-resourced personnel agent. This is a deliberately stressed parameterization designed to exercise the variance dynamics; its purpose is to make the degraded-controls regime visible, not to estimate real-world hospital ransomware incidence.

\paragraph{Scenario B: Medium controls.} 18 LECs including SIEM-based detection, resistance LECs at 70--100\% intended efficacy, fast VMC detection (2--4 hour sweeps), strong DSCs (90--100\%), one well-resourced personnel agent. This represents an adequately resourced organization where controls are well-configured and monitoring is responsive.

Unless otherwise stated, reported numbers are from the medium-controls variant.

\section{Core Mechanics}

\subsection{Susceptibility formula (breach probability)}

Per the FAIR-CAM standard, combined susceptibility uses the defense-in-depth pass-through formula. Each resistance LEC protecting a target asset is modeled as an independent hurdle that a threat must pass through:
\begin{equation}
    S_{\text{combined}} = \prod_{i} (1 - \mathrm{OpEff}_i) \quad \text{for all resistance LECs protecting the target}
    \label{eq:suscept-detail}
\end{equation}
Combined Resistance Strength is $\mathrm{RS} = 1 - S_{\text{combined}}$. A breach occurs when the threat agent's sampled sophistication (TCap) exceeds RS:
\[
    \text{breach} \iff \mathrm{TCap} > \mathrm{RS}
\]

Stochasticity enters through TCap, which is sampled fresh from the threat agent's Beta-PERT distribution at each contact event. Resistance control efficacies are set-and-hold --- sampled once at initialization and after each remediation, held constant between state transitions. The comparison itself is deterministic: given a sampled TCap and the current RS, the outcome is binary.

Properties: adding controls always increases RS (reducing breach probability); zero controls means $\mathrm{RS} = 0$ (certain breach); any control with $\mathrm{OpEff} = 1.0$ makes $\mathrm{RS} = 1.0$ (breach impossible). Threat sophistication also affects pre-breach stages: avoidance and deterrence LECs compare their efficacy against TCap to determine whether the threat is blocked before reaching the resistance stage.

\subsection{Detection: two distinct planes}
\label{sec:detect}

The simulator implements detection at two structurally distinct planes that are easily conflated and which the FAIR-CAM standard treats separately.

\paragraph{LEC detection (breach detection, V$\wedge$M$\wedge$R AND-gate).} When a threat event reaches the asset, detection of the resulting breach requires all three loss-event-control subfunctions to be operational: visibility (making threat activity observable), monitoring (actively looking for it), and recognition (correctly interpreting what is observed). Failure of any single subfunction yields complete detection failure:
\begin{equation}
    P(\text{detect breach}) = \mathbb{1}[V_{\text{op}} \wedge M_{\text{op}} \wedge R_{\text{op}}] \cdot p_{V} \cdot p_{M} \cdot p_{R}
    \label{eq:detect-and}
\end{equation}
where the indicator is $1$ iff each subfunction has at least one operational LEC and $p_{V}, p_{M}, p_{R}$ are the per-event Bernoulli success probabilities of those operational LECs (derived from their efficacies). If the indicator evaluates to $0$, detection fails and the breach goes undetected for a configurable dwell time (default 264 hours, calibrated to the M-Trends 2025 global median dwell time of 11 days~\citep{MTrends2025}). Undetected breaches bypass the entire response pipeline.

\paragraph{VMC detection (variance detection, monolithic in this version).} A separate, distinct detection plane governs whether the organization notices that a control has drifted out of its intended state. The simulator currently implements this as a monolithic process: each VMC samples a detection interval (Beta-PERT) and at each sweep either detects all variant controls it monitors with a configured probability, or detects none. The full FAIR-CAM stage-gated decomposition of VMC detection into Visibility / Monitoring / Recognition sub-functions is not implemented in this version; this is an explicit simplification (see Section~\ref{sec:knownlim}). The two planes interact only indirectly: a variant detection LEC degrades the V$\wedge$M$\wedge$R AND-gate above and is itself subject to discovery by a VMC monitoring sweep (Section~\ref{sec:cascade}).

\subsection{Loss magnitude}

Loss magnitude depends on the type of business asset compromised. For information assets, gross loss is sampled from empirically calibrated lognormal distributions parameterized by sector and scenario type, with parameters derived from the Cyentia IRIS 2025 report~\citep{IRIS2025}. For process assets, loss is driven by outage duration (the sum of detection time, containment time, and recovery time) with lookup from outage cost tables anchored on NetDiligence business interruption claims data (5-year average, SME ransomware, $N=294$ claims)~\citep{NetDiligence2024} and scaled by duration bracket.

Total loss is split into primary and secondary loss. The current implementation does not yet decompose either side into the six FAIR forms of loss (productivity, response, replacement, fines \& judgments, reputation, competitive advantage) individually. Response and loss-reduction LECs reduce gross loss to net loss, but only when the breach was detected (Section~\ref{sec:detect}).

\subsection{Calibration sources}
\label{sec:dataprovenance}

The model's empirical parameters are drawn from three sources: the Cyentia Institute IRIS 2025 report~\citep{IRIS2025} (insurance claims, $N{>}150{,}000$ events) for loss magnitude distributions, the NetDiligence Cyber Claims Study~\citep{NetDiligence2024} (insurance claims, $N{=}294$ ransomware BI claims) for business interruption cost anchors, and Mandiant M-Trends 2025~\citep{MTrends2025} (incident response telemetry) for dwell time calibration. All three are insurance claims or incident response data; their known population biases (insured organizations only for IRIS and NetDiligence, externally investigated incidents for M-Trends) propagate into the simulation's loss magnitudes and timing but do not affect the organizational dynamics that are the paper's focus.

\subsection{Human-actor controls and DSC integration}

The simulator implements a DSC alignment model for human-operated controls: when a LEC's actor type is ``human,'' its efficacy check is mediated by a five-dimension personnel alignment assessment (expectation, awareness, capability, situational awareness, incentive) rather than a single Bernoulli trial. Misalignment degrades the control's effective efficacy proportionally. This mechanism is disabled in all experiments reported here; its activation and calibration are reserved for a dedicated follow-on study.

\section{Variance Model}
\label{sec:variance}

\subsection{Three sources of variance}

The FAIR-CAM standard identifies three sources of control variance, each modeled as a distinct mechanism in the simulation.

\paragraph{Intrinsic regular (control drift).} Each control samples a \texttt{next\_change\_time} from its Change Frequency distribution (Beta-PERT). When reached, variance is attempted, gated by DSC alignment and VMC ``reduce variance probability'' checks. Models natural degradation: patches go stale, configurations drift.

\paragraph{Intrinsic irregular (personnel-driven).} Monthly assessment: personnel with admin privileges who are misaligned (per DSC model) can degrade the controls they manage. DSC and VMC controls can prevent this.

\paragraph{Extrinsic (threat landscape).} Poisson process (configurable, default of ${\sim}1$/year for the experiments reported in this paper). All software-based controls instantly become variant. Bypasses all prevention gates. Models zero-days, new CVEs. We note that this is a deliberate worst-case simplification: in reality a zero-day or new CVE typically affects a specific subset of software controls (those exposed to the new vector), not all software controls simultaneously. The simulator's all-software-LECs-at-once formulation provides an upper bound on extrinsic correlation effects; per-control extrinsic susceptibility (e.g., conditioning on technology stack) is left to future work.

\subsection{Variant efficacy}

When a control enters variant state, its operational efficacy degrades. The magnitude of degradation depends on whether the control is binary or continuous:
\begin{align}
    \text{Non-binary:} \quad \mathrm{VarEff} &= \mathrm{Uniform}(0, \mathrm{IntEff}) \\
    \text{Binary:} \quad \mathrm{VarEff} &= 0.0
\end{align}

\subsection{Recovery path}

The recovery path from variant state to normal follows a multi-stage pipeline. A variant control must first be \emph{detected} by a VMC monitoring sweep --- controls do not self-heal. Once detected, the control enters the remediation queue, where it waits for available budget capacity. The queue is priority-ordered by control type (resistance $>$ detection $>$ response $>$ resilience), then by scheduling strategy, then FIFO:
\begin{multline*}
    \text{Variant} \to \text{VMC detection sweep} \to \text{Remediation queue (budget-gated)} \\
    \to \text{Remediation complete} \to \text{Normal}
\end{multline*}

On return to normal, IntEff is re-sampled from the Beta-PERT distribution.

\subsection{Budget-constrained remediation}

The remediation queue implements finite organizational capacity. The default budget is 40 hours per month --- roughly one FTE at 25\% allocation to security remediation. The budget cadence is configurable (monthly reset, quarterly pooling, or continuous accrual). Within each budget period, the queue starts items in priority order until budget is exhausted.

Remediation is not instantaneous. The total time a control spends out of its intended state has four additive components, all of which the simulator tracks explicitly:
\begin{equation}
    \text{time in variant} = t_{\text{detect}} + t_{\text{queue wait}} + t_{\text{remediation}} + t_{\text{backtrack}}
    \label{eq:timein}
\end{equation}
where $t_{\text{detect}}$ is the delay between the variance event and the next VMC monitoring sweep that flags the control; $t_{\text{queue wait}}$ is the time the control sits in the priority queue after detection but before remediation begins (zero if budget is non-binding, positive otherwise); $t_{\text{remediation}}$ is the per-control duration drawn from configuration parameters or defaults (8 hours by default, with $4 / 16 / 40$ hours for low / medium / high-severity remediation classes); and $t_{\text{backtrack}}$ is an optional culture-modulated delay. The remediation duration is further scaled by a Competing Values Framework (CVF) culture modifier reflecting the organization's responsiveness profile, and may be gated by a separate VMC \emph{implements} relationship: if the VMC responsible for performing the remediation is itself variant, remediation cannot start. Together these terms produce the sawtooth queue-depth pattern visible in the backlog dynamics figure (Figure~\ref{fig:backlog}).

The remediation pipeline is structurally a single-server queueing system: variance events arrive (according to a process that is approximately Poisson at the level of intrinsic drift plus extrinsic shocks), enter a priority queue, and are served at a rate bounded by the monthly engineer-hour budget. Classical queueing theory predicts the qualitative behaviour of such a system as utilization $\rho = \lambda / \mu$ (arrival rate over service rate) varies~\citep{Little1961, Kleinrock1975}: when $\rho$ is comfortably below 1, the queue empties between arrivals and waiting time is short; when $\rho$ approaches 1, expected wait time grows non-linearly; when $\rho \geq 1$, the queue grows without bound and outcomes are dominated by the priority discipline rather than by the budget magnitude.

In the FAIR-CAM setting, the cyber-risk-relevant consequence of a saturated queue is that, while controls wait, each unrepaired variant control raises combined susceptibility (Equation~\ref{eq:suscept}). The analytical OpEff formula cannot capture this because it assumes each control's reliability is independent and stationary, with no notion of a shared service capacity. The simulation makes the queueing regime observable end-to-end and locates the budget threshold at which the system transitions from non-binding (small $\rho$) to saturated (large $\rho$) for a given control portfolio and threat-arrival rate.

We deliberately label this an \emph{accumulation under a capacity constraint}, not a closed reinforcing loop: breaches consume the loss pipeline but do not modify upstream variance arrivals or budget allocation (Section~\ref{sec:knownlim}).

All experiments in this paper use the monthly-reset budget and the default culture profile; sensitivity to cadence and culture is deferred to a follow-on paper.

\section{Organizational Dynamics}

\begin{figure}[htbp]
\centering
\includegraphics[width=0.95\textwidth]{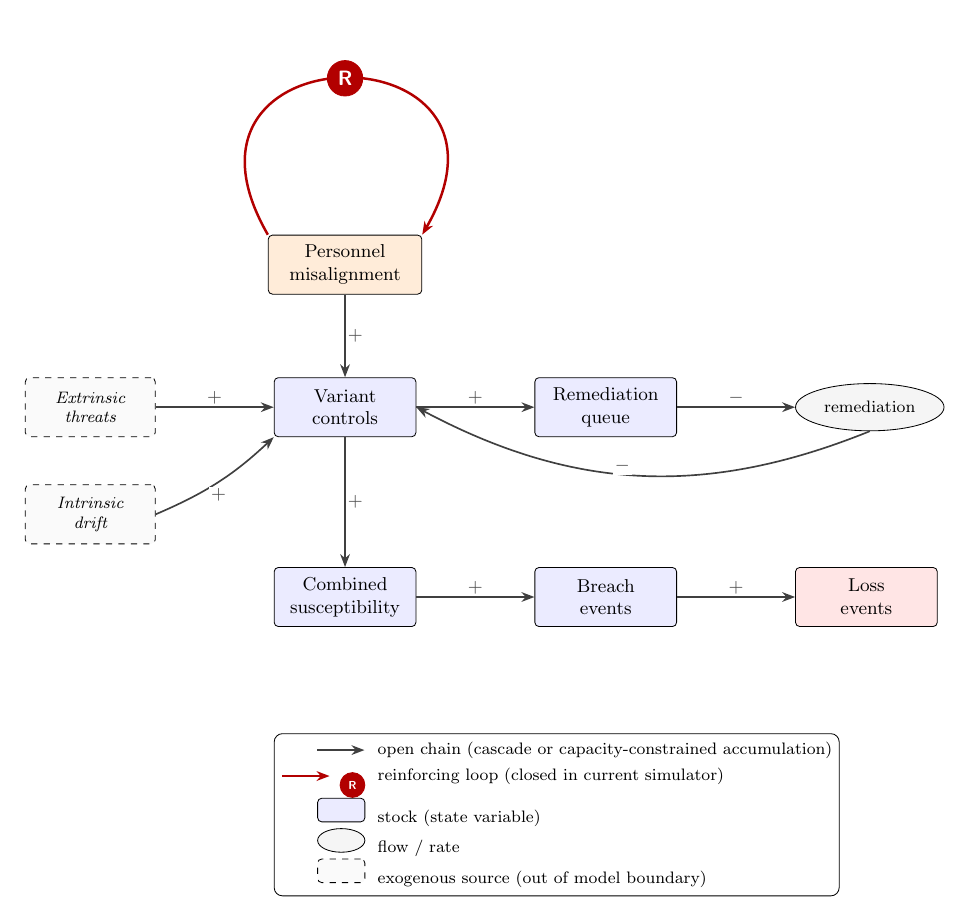}
\caption{Causal structure of the simulator. Solid black arrows: open chains forming cascades (variance $\to$ susceptibility $\to$ breach $\to$ loss) and a capacity-constrained accumulation (variant controls $\to$ remediation queue $\to$ remediation $\to$ variant controls). The red R loop (personnel misalignment contagion) is the only closed reinforcing loop in the simulator; it is disabled in all experiments reported here. Three arcs that would close the open chains into feedback loops (Loss $\to$ awareness, Loss $\to$ budget, Loss $\to$ intrinsic variance) are identified in Section~\ref{sec:discussion} as future work.}
\label{fig:cld}
\end{figure}

This section presents three organizational dynamics that the simulator makes observable. We deliberately distinguish their causal structures: Section~\ref{sec:opeffdiv} shows that simulated operational efficacy diverges from analytical values. Section~\ref{sec:budgetregime} shows that the remediation pipeline accumulates unresolved controls under capacity constraints (a queueing regime shift). Section~\ref{sec:cascade} shows that failures propagate through the monitoring topology (one-way dependencies). The experiments exercise no closed feedback loops; the arcs that would close them are identified in Section~\ref{sec:discussion}. Figure~\ref{fig:cld} summarizes the distinction.

\subsection{The formula overestimates under realistic conditions}
\label{sec:opeffdiv}

The FAIR-CAM analytical formula defines operational efficacy as Equation~\eqref{eq:opeff}, where each parameter is treated as a time-invariant property of the control. Under stationary conditions (constant change frequency, remediation budget that exceeds demand, and reliable VMCs) the formula is correct and sufficient. In the simulation, however, OpEff is not computed from inputs but emerges from the state machine dynamics: reliability is the empirical fraction of time a control spends in its intended state over the simulation run.

This convergence under stationarity serves as an implementation-consistency check: a confirmation that the state-machine correctly reproduces the analytical formula in the regime for which that formula is derived. Under non-stationary conditions, the ABM diverges from the analytical formula. The divergence is not a rejection of the formula; it quantifies where the formula's stationarity assumption ceases to hold. Three organizational mechanisms drive the divergence:
\begin{enumerate}[nosep]
    \item \textbf{Budget-constrained remediation} creates temporal correlations: when remediation capacity is saturated, multiple controls remain variant simultaneously, producing correlated degradation. This is the security-specific instance of the ``compliance budget'' phenomenon identified by \citet{Beautement2008}: organizations have finite capacity for security-related work, and exceeding that capacity forces prioritization trade-offs.
    \item \textbf{VMC cascading failures} create hidden state: when a VMC becomes variant, the controls it monitors accumulate undetected variance.
    \item \textbf{Personnel-driven variance} introduces organizational correlation: social influence and contagion create clusters of misaligned personnel who simultaneously degrade multiple controls.
\end{enumerate}

\begin{figure}[htbp]
\centering
\includegraphics[width=0.85\textwidth]{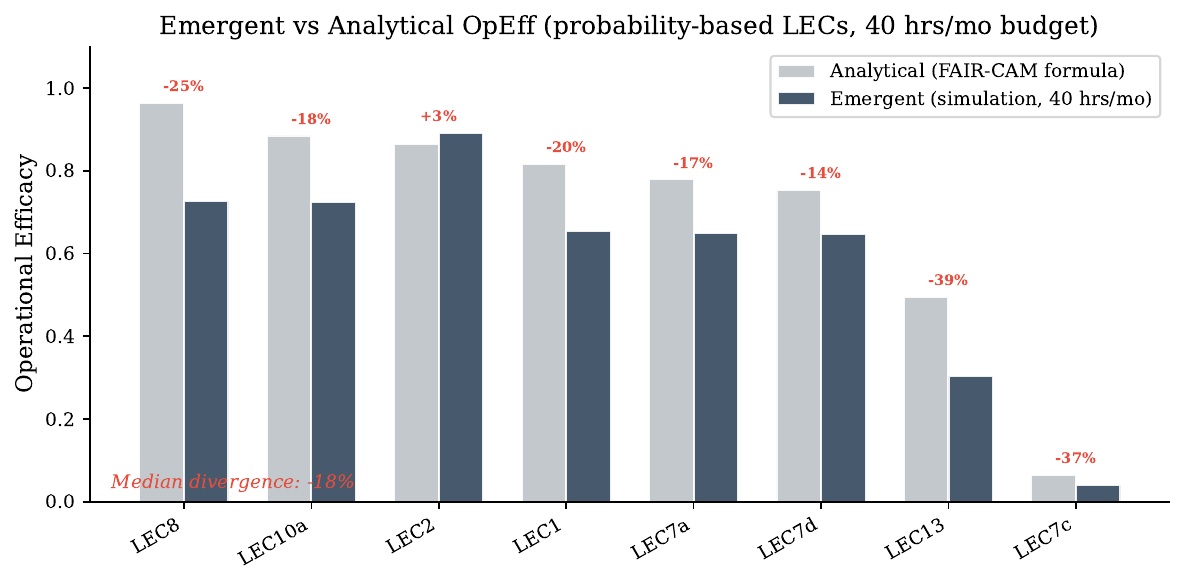}
\caption{Emergent vs analytical operational efficacy for the 8 highest-efficacy probability-based LECs at budget $=40$ hr/mo (1{,}000 iterations, 1-year horizon). Light bars: analytical FAIR-CAM value. Dark bars: emergent mean reliability from the simulation. Per-LEC divergence (emergent $-$ analytical, percentage points) is annotated above each pair; the median across all 14 fractional-efficacy LECs is $-17.4\%$ and is essentially flat across budget levels. A complementary threat-landscape frequency sweep (Table~\ref{tab:threat-freq}) gives $-16.8\%$ at baseline frequency and vanishes at zero extrinsic frequency, confirming correlated extrinsic variance as the primary driver.}
\label{fig:opeff-divergence}
\end{figure}

\paragraph{Quantifying the divergence.} We report divergence from two complementary experiments. In the \emph{budget sweep} (varying the remediation-budget factor at 7 levels from 2 to 160 hours/month, 1{,}000 iterations per level, baseline extrinsic frequency of 1 event/year), emergent OpEff for fractional-efficacy LEC controls is systematically lower than the analytical value. Aggregating per-control mean emergent OpEff across 14 LECs with analytical values in $(0, 1)$, the median per-LEC divergence is $-17.4\%$, with individual LECs ranging from $-38.6\%$ to $+3.3\%$ (Figure~\ref{fig:opeff-divergence}). In the \emph{frequency sweep} (1{,}000 iterations at fixed budget $=40$ hrs/mo, varying extrinsic frequency from 0 to 1), the median divergence at the same baseline frequency is $-16.8\%$. The two independent sweeps agree in direction, sign, and approximate magnitude. A control with analytical OpEff of 0.80 therefore has an emergent value of approximately 0.64--0.67 --- a gap that compounds through the multiplicative susceptibility formula.

The divergence is not budget-driven. The median per-LEC divergence is essentially identical across budget levels from 2 to 160 hrs/mo. We ran an ablation in which the recovery-time resampling of \texttt{IntEff} from its Beta-PERT distribution (Section~\ref{sec:variance}) was disabled, to test whether a Jensen-style numerical bias from the resampling was responsible for the divergence. Disabling resampling does not reduce the divergence. The resampling is therefore not the driver.

\paragraph{Extrinsic frequency is the primary driver.} As a factor-prioritization exercise~\citep{Saltelli2008} --- identifying which input most drives the divergence --- we varied the extrinsic-frequency factor at 11 equally-spaced levels from 0 to 1 events/year, holding budget fixed at 40 hours/month, 1{,}000 iterations per level (seeds 0--999). This sweep and the budget sweep above test distinct hypotheses (budget sensitivity and extrinsic-frequency sensitivity); no multiple-comparison correction is applied because the sweeps are independent experiments, each with a single pre-specified direction of effect (these are one-at-a-time sensitivity exercises in the sense of Saltelli et al.~\citep{Saltelli2008}, not variance-based global sensitivity analyses). Table~\ref{tab:threat-freq} reports the results.

\begin{table}[htbp]
\centering
\caption{Threat landscape frequency sweep (medium controls, 1{,}000 iterations, budget = 40 hrs/mo). Median per-LEC divergence (emergent $-$ analytical) as a function of extrinsic threat landscape change frequency.}
\label{tab:threat-freq}
\begin{tabular}{rrrr}
\toprule
Freq/year & Median div. (\%) & Mean ALE (\$) & Mean breaches \\
\midrule
0.0 & $+0.1$  & 0       & 0.000 \\
0.1 & $-0.9$  & 28{,}527  & 0.000 \\
0.2 & $-3.3$  & 97{,}274 & 0.020 \\
0.3 & $-5.4$  & 3{,}925{,}891 & 0.050 \\
0.4 & $-7.7$ & 657{,}021 & 0.050 \\
0.5 & $-9.9$ & 300{,}045 & 0.030 \\
0.6 & $-11.2$ & 426{,}070 & 0.050 \\
0.7 & $-12.1$ & 920{,}643 & 0.080 \\
0.8 & $-15.9$ & 1{,}434{,}391 & 0.120 \\
0.9 & $-15.7$ & 845{,}973 & 0.080 \\
1.0 & $-16.8$ & 930{,}875 & 0.090 \\
\bottomrule
\end{tabular}
\end{table}

At zero extrinsic frequency, where the only variance source is each control's intrinsic drift timer, emergent OpEff converges to the analytical formula (median divergence $+0.1\%$). As extrinsic frequency increases, the median divergence grows approximately linearly at ${\sim}1.80$ percentage points per 0.1 events/year. At the scenario's baseline frequency of 1 event/year, the median divergence reaches $-16.8\%$ (95\% bootstrap CI on the median: $[-19.3\%,\ -2.5\%]$). Note that the mean ALE column is non-monotonic because loss events are rare at $N=1{,}000$ under the medium scenario (the breach rate is only $\sim$3\% of iterations at baseline $\rho$); the ALE estimator has high variance and should not be interpreted as a smooth function of extrinsic frequency. The median divergence, computed across all fractional-efficacy LECs, is the stable signal.

The mechanism is the batch-arrival effect: a threat landscape change simultaneously pushes software-based controls into variant state, creating a remediation batch that exceeds the queue's monthly throughput capacity. Under purely intrinsic drift, variance events arrive staggered according to each control's independent change-frequency distribution (in the tested scenario typically Uniform(25{,}000--45{,}000) hours $\approx$ 2.9--5.1 years between events), and the 40-hour monthly budget clears each one before the next arrives. Under correlated extrinsic shocks, the same budget must service the entire portfolio simultaneously, producing the queue saturation and multi-control degradation that the stationary analytical formula cannot represent. The specific magnitude is scenario-dependent, but the qualitative point generalises: any organization facing correlated extrinsic variance (zero-days, regulatory changes, supply-chain events) should expect the analytical OpEff formula to overestimate by a margin proportional to extrinsic event frequency.

\subsection{Budget creates a queueing regime shift}
\label{sec:budgetregime}

Remediation budget does not reduce risk linearly. Below a critical threshold, the system enters a degraded regime where the remediation queue cannot clear between extrinsic variance events. Above the threshold, additional budget has zero marginal value.

\begin{figure}[htbp]
\centering
\includegraphics[width=0.85\textwidth]{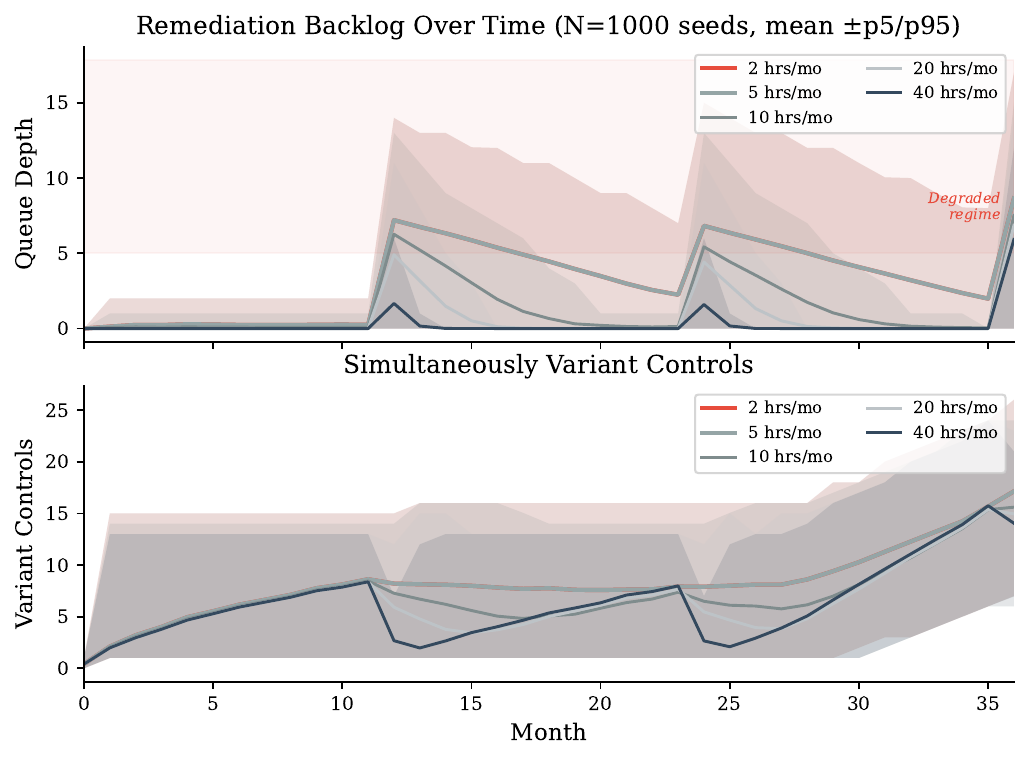}
\caption{Ensemble 3-year trajectories across $N=1{,}000$ seeds at five budget levels. Top: mean remediation queue depth (solid lines) with 5th--95th percentile shaded band per budget over 36 months; red band marks the degraded regime (queue depth $>5$). Bottom: mean simultaneously variant controls with p5--p95 band. Extrinsic event arrivals are Poisson-distributed per seed (exponential inter-arrival with mean 1/year); at the scenario's baseline frequency the ensemble-averaged trajectory produces clustered reactivity peaks around months ${\sim}11$--12 and ${\sim}23$--24, visible across all budgets (the per-seed p5--p95 bands absorb the inter-arrival dispersion). At 2--5 hrs/mo the queue does not clear between these events and controls accumulate in variant state; at 10+ hrs/mo the queue drains within each inter-event interval. Traces for 2 and 5 hrs/mo are visually coincident because at both budgets the per-cycle arrivals exceed throughput, producing identical steady-state dynamics.}
\label{fig:backlog}
\end{figure}

In 3-year ensemble simulations across $N=1{,}000$ seeds (0--999) at five budget levels, the transition is sharp. At 2--5 hrs/mo, the remediation queue maintains a mean time-averaged depth of 3.2 controls (5th--95th percentile across seeds: 0.4--6.6) --- a persistent backlog that cannot clear before the next extrinsic event arrives; 8.2 of 37 controls are simultaneously variant on average (p5--p95: 3.0--12.5). Mean breach count rises from 0.54 at 40 hrs/mo to 1.13 at 2 hrs/mo per 3-year run (+111\%). At 10 hrs/mo, the queue clears within each budget cycle (mean depth 1.4, 0.58 breaches/3yr). At 20--40 hrs/mo, mean queue depth drops to 0.3--0.7 and breach counts converge (0.54--0.56/3yr), confirming that additional budget beyond ${\approx}10$ hrs/mo has diminishing marginal risk-reduction value in this scenario (Figure~\ref{fig:backlog}). The Section~\ref{sec:validation} OpEff sweep at 80 and 160 hrs/mo corroborates the saturation.

The mechanism is a race between variance arrival and remediation throughput (Section~\ref{sec:variance}): an extrinsic event creates a remediation batch; if the budget cannot clear it before the next event, the residual compounds.

\subsection{Monitoring cascades}
\label{sec:cascade}

Budget constraints determine how long variant controls wait for repair; monitoring constraints determine whether they are discovered at all. The most severe organizational dynamic occurs when VMCs themselves degrade. A variant VMC cannot detect variance in the controls it manages. Those controls accumulate undetected degradation, never entering the remediation queue, while the organization's monitoring infrastructure reports a healthy posture.

The cascade deepens when VMCs manage other VMCs. In the hospital scenarios, VM4 (vulnerability scanning) monitors several LECs, while VM2 (threat monitoring) provides threat intelligence that reduces detection time. When VM4 is variant, the LECs it monitors accumulate undetected variance. When VM2 is simultaneously variant, detection time increases, meaning any breach that does occur produces higher losses due to longer dwell time.

\begin{figure}[htbp]
\centering
\includegraphics[width=0.85\textwidth]{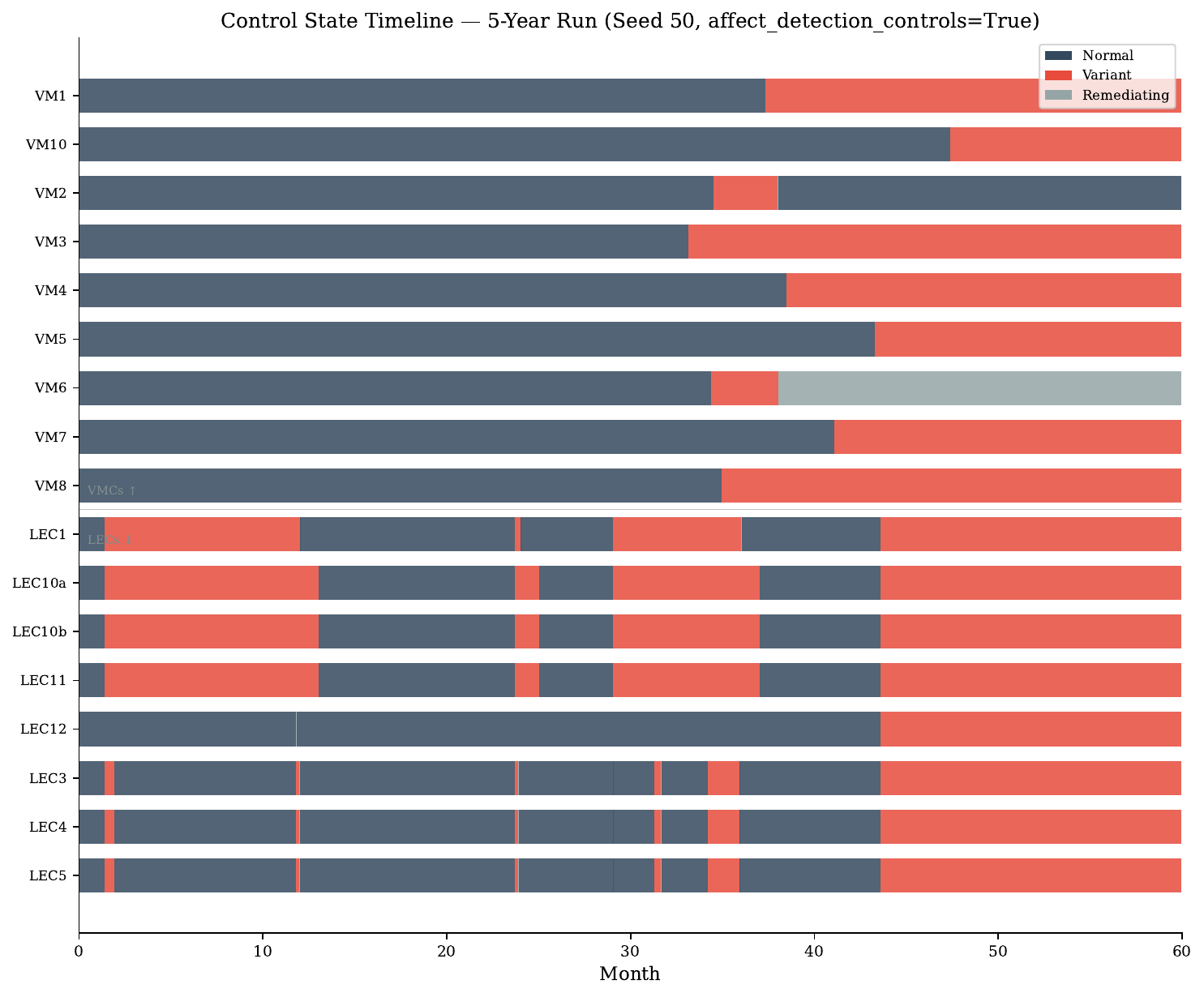}
\caption{Per-control state timeline over 5 years (seed 50, \texttt{affect\_\allowbreak detection\_\allowbreak controls\allowbreak =True}). VMCs above the separator, cascade-involved LECs below. Navy = normal, rust = variant, grey = remediating. From month 33, VM3's drift initiates a cascade through the monitoring hierarchy.}
\label{fig:cascade}
\end{figure}

In an illustrative 5-year simulation (seed 50, \texttt{affect\_\allowbreak detection\_\allowbreak controls\allowbreak =True}) with threat landscape events degrading VMCs alongside LECs, the narrative engine identified three cascade windows totaling over 53,000 hours of compounding detection failure. The most severe: VM4 went variant at month 38; 17 LECs and 2 additional VMCs accumulated undetected variance over 15,672 hours. A VMC-to-VMC chain --- VM3 $\to$ VM4 $\to$ VM7 $\to$ VM5 --- demonstrated propagation through the monitoring hierarchy: by month 43, four of nine VMCs were simultaneously variant, monitoring approximately 70\% of the LEC population without the ability to detect their degradation (Figure~\ref{fig:cascade}). The specific figures should be read as the representative trajectory of a single high-breach seed; the distribution across seeds is quantified below.

Across an ensemble of $N=1{,}000$ seeds under identical configuration, 90.5\% of seeds exhibit at least one VMC cascade window and 23.4\% exhibit three or more. Cascade durations are long-tailed: median 50.9 months (p5--p95: 19.7--59.7 months). Three VMCs --- VM4, VM3, and VM8 --- account for all observed cascades in this scenario (59\%, 21\%, 20\% of windows respectively); they are structural linchpins whose degradation allows variance to accumulate undetected in the LECs they monitor. Seeds experiencing at least one cascade produce $1.95\times$ more breaches (34.0 vs 17.4 per 5-year run) than cascade-free seeds, confirming that the cascade mechanism is not a rare outlier but a near-universal dynamic in this scenario.

This cascade pattern is uniquely an organizational dynamic, consistent with Cook's observation that complex system failures arise from the combination of latent conditions rather than single-point causes \citep{Cook2000}. The simulation reveals that a single VMC failure can propagate through the monitoring topology, producing correlated, invisible degradation across the control ecosystem. For practitioners, this argues for VMC redundancy and independent monitoring paths.

\paragraph{Three paths to LEC variance: a complete VMC cascade taxonomy.}
The cascade illustrated above traces a single VMC failure mode, detection failure. The full FAIR-CAM standard distinguishes four VMC sub-functions (variance prevention by reducing change frequency, variance prevention by reducing variance probability, variance detection, and variance correction), which collapse into three distinct paths by which a degraded VMC contributes to an LEC entering and remaining in the variant state:
\begin{enumerate}[nosep]
    \item \textbf{Prevention failure.} Variance prevention is an \emph{active} VMC function: when an LEC's internal drift timer fires, the simulator first consults any VMCs linked to that LEC whose role is to reduce variance frequency or variance probability. These prevention VMCs run a Bernoulli check against their current efficacy and, on success, block the variance event before it is applied --- the LEC stays in its intended state and its drift timer is reset. If the prevention VMC is itself in a variant state at the moment the LEC's timer fires, its efficacy is degraded and it is correspondingly less likely to block the drift. The net effect is that the LEC enters the variant state on drift events it would have survived under a healthy prevention VMC. Cascade-tracing in this path therefore asks, for each variance event on an LEC, whether the VMCs responsible for preventing it were themselves healthy at the moment the LEC drifted.
    \item \textbf{Detection failure.} A VMC whose function is to monitor an LEC for drift is variant during the LEC's variant interval, so the variance event is not enqueued for remediation. The LEC stays variant longer than it would have under a healthy detection VMC. This is the path the cascade narrative above traces.
    \item \textbf{Correction failure.} A VMC whose function is to implement the remediation is variant when the LEC reaches the front of the queue, so remediation cannot start even though the budget would have permitted it. The LEC stays variant until the correction VMC itself recovers.
\end{enumerate}
The simulator implements all three failure paths in the variance and remediation pipelines (with prevention failure handled at variance attempt, detection failure at the VMC monitoring sweep, and correction failure at remediation start). The narrative engine and linchpin analysis (Section~\ref{sec:rca}) trace all four cascade paths: detection (monitoring VMC variant during LEC variant interval), prevention (variance-reduction VMC variant at LEC drift time), threat intelligence (intel VMC variant at breach time, increasing detection latency), and correction (implementing or treatment-selection VMC variant, blocking or deprioritizing remediation). Each path uses temporal overlap checking with recovery events to eliminate false positives --- a VMC that was variant but recovered before the causally relevant tick is not counted.

\section{Narrative Causation Engine}

\subsection{Design}

The narrative causation engine records every significant event during the simulation: variance events (which control degraded, why, when), breach events (which threat, which asset, which controls were variant), detection events (which VMC detected what, or failed to), and loss events (magnitude, detection status, response outcome). These events are linked by causal relationships: a loss event references the breach that caused it, which references the variant controls that enabled it, which reference the variance events that degraded them.

For each loss event, the engine produces a causation chain: the specific controls that were variant at breach time, why each was variant, which VMCs contributed through any of the four cascade paths (Section~\ref{sec:cascade}), and which personnel decisions allowed the variance to persist.

Linchpin analysis identifies the controls whose variance contributed most to loss events across four cascade paths. For each loss event, the engine checks whether linked VMCs were variant at the causally relevant moment: at the LEC's drift time for prevention VMCs (variance-probability and change-frequency reduction), during the LEC's variant interval for detection VMCs (controls monitoring), at breach time for threat-intelligence VMCs (whose variance increases detection latency), and during the variant-to-remediation window for correction VMCs (implementing or treatment-selection VMCs whose variance blocked or deprioritized remediation). Each breach is classified by root cause using counterfactual analysis. When resistance controls were variant at breach time, the engine recomputes the combined resistance strength from intended efficacies and compares it against the threat's sophistication: if the threat would have breached even at full intended efficacy, the classification is \emph{threat\_exceeded} (variance was coincidental); if the threat would have been resisted at intended efficacy, the classification is \emph{variance\_enabled} (variance was causally necessary for the breach). The third category, \emph{missing\_controls}, applies when no LECs existed in the protection path.

\subsection{Root cause analysis}
\label{sec:rca}

Loss events are further classified by loss driver category, mapping to the RCA dimensions defined in the FAIR-CAM standard:
\begin{itemize}[nosep]
    \item \emph{Detection failure} --- controls operational but the detection pipeline failed (suggests VMC investment)
    \item \emph{Resource constrained} --- controls in remediation queue but budget exhausted (suggests budget increase)
    \item \emph{Control degradation} --- controls drifted from intended state (suggests monitoring improvement)
    \item \emph{Design weakness} --- architecture insufficient at full efficacy (suggests redesign)
\end{itemize}

The loss driver classification directly informs investment decisions: if 70\% of losses are resource-constrained, increasing the remediation budget is the highest-leverage intervention; if 70\% are detection failures, investing in VMC reliability matters more.

\paragraph{Completeness verification.} We verified trace completeness across both hospital scenarios (100 iterations medium, 30 iterations weak, 806 total loss events). For each loss event we checked nine properties: presence of event ID, breach tick, positive total loss, classified breach category (\texttt{variance\_\allowbreak enabled} $|$ \texttt{threat\_\allowbreak exceeded} $|$ \texttt{missing\_\allowbreak controls}), classified loss driver, non-empty causation narrative text, populated breach mechanics, identified failed controls, and resolved root causes. All 806 loss events passed all nine checks (100\% completeness). This confirms that the engine produces a complete causal trace for every loss event the simulation emits, without exception.

\section{Results and Convergence Check}
\label{sec:validation}

This section reports a budget sensitivity analysis and scenario-level results. All experiments use $N=1{,}000$ iterations; convergence testing confirmed that the coefficient of variation of mean ALE stabilizes below 5\% at $N \geq 200$ (convergence plots available in the repository).

\subsection{Budget sensitivity analysis}

We performed a budget sensitivity sweep using the medium controls scenario (1{,}000 iterations per budget level, 1-year horizon, seeds 0--999). Remediation budget was varied as a single factor from 2 to 160 hours/month, holding all other parameters constant (a one-factor-at-a-time design~\citep{Montgomery2017}). Results are shown in Table~\ref{tab:budget}.

\begin{table}[htbp]
\centering
\caption{Budget sensitivity analysis (medium controls, 1{,}000 iterations, 1-year horizon, seeds 0--999). Rows with statistically equivalent results are collapsed into regime bands (see footnote).}
\label{tab:budget}
\footnotesize
\begin{tabular}{l r r r r}
\toprule
\textbf{Budget (hr/mo)} & \textbf{Mean ALE} & \textbf{P95 ALE} & \textbf{Breaches} & \textbf{Breach iters} \\
\midrule
$\leq 5$ (saturated)       & \$2{,}586{,}392 & \$4{,}856{,}303 & 254 & 7.2\% \\
10                          & \$1{,}103{,}605  & \$0         & 96  & 4.0\% \\
20                          & \$1{,}026{,}970  & \$0         & 94  & 3.6\% \\
$\geq 40$ (non-binding)    & \$930{,}875  & \$0         & 85  & 3.3\% \\
\bottomrule
\end{tabular}
\\[0.5em]
\footnotesize\raggedright
\textit{Note.} Budget levels within each regime produce statistically equivalent results across 1{,}000 paired seeds. P95~$=\$0$ for budgets $\geq 10$ reflects the heavy-tailed loss distribution: only 3.3--4.0\% of iterations produce any breach, so the 95th percentile falls within the zero-loss mass (P99 is \$24.5--29.9M).
\end{table}

\begin{figure}[htbp]
\centering
\includegraphics[width=0.85\textwidth]{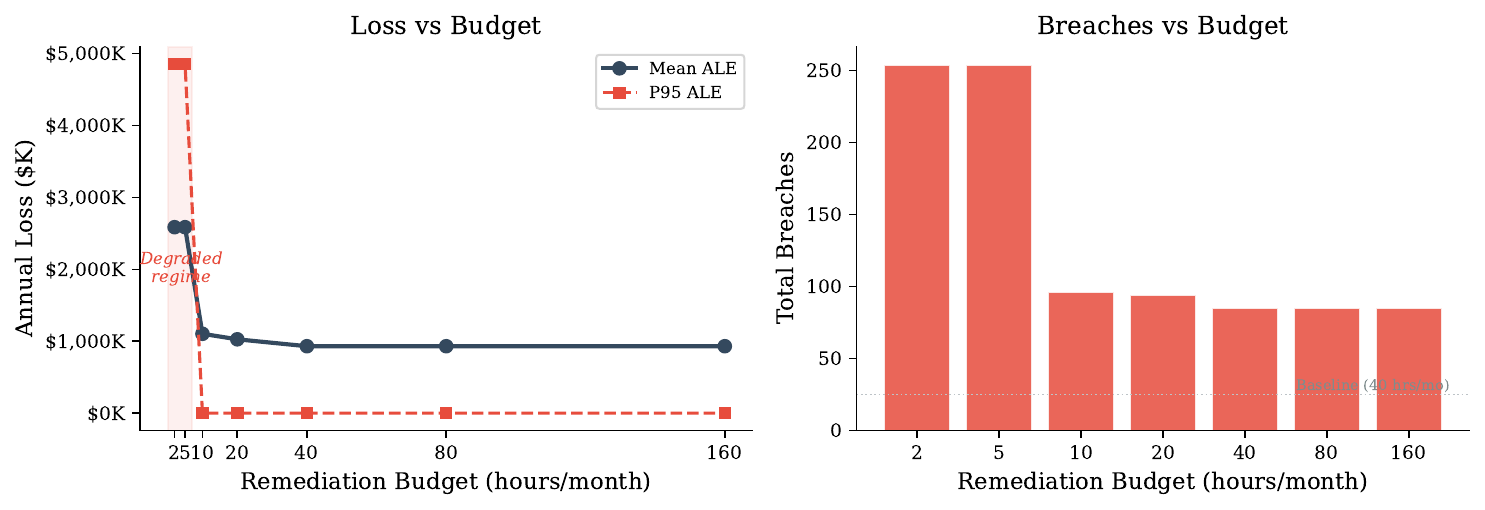}
\caption{Budget sensitivity analysis for the medium hospital ransomware scenario (1{,}000 iterations, 1-year horizon). Left panel: mean ALE and P95 ALE as a function of remediation budget. Right panel: total breaches across 1{,}000 iterations by budget level. A regime shift occurs between 5 and 10 hours/month, bounded by a saturated band ($\leq 5$, mean ALE \$2.59M) and a non-binding band ($\geq 40$, mean ALE \$931K). The location and width of the transition are scenario-specific (see Section~\ref{sec:budgetregime}).}
\label{fig:budget}
\end{figure}

The results reveal three distinct queueing regimes (Figure~\ref{fig:budget}) corresponding to the classical utilization map of a single-server priority queue~\citep{Little1961, Kleinrock1975}.

\textit{(i) Saturated regime} ($\leq 5$ hr/mo, $\rho \gg 1$). Budget capacity is far below remediation demand. The queue is full in every cycle and the order of remediation is determined entirely by the priority discipline (resistance $>$ detection $>$ response $>$ resilience, then FIFO), independent of the budget magnitude. Budget$=2$ and budget$=5$ produce statistically equivalent results across 1{,}000 seeds (identical mean ALE, identical breach count).

\textit{(ii) Binding regime} (5--20 hr/mo, $\rho \approx 1$). Capacity is comparable to demand. Whether the queue clears between successive variance events --- and which controls do or do not finish remediation in time --- depends on the exact budget value, and small budget changes produce real outcome differences. The key paired-seed comparison sits at the saturated-to-binding boundary: reducing the budget from 10 to 5 hours/month (i.e.\ pushing the system from $\rho \approx 1$ into $\rho \gg 1$) increases mean ALE by \$1{,}482{,}787 (95\% paired-bootstrap CI $[\$143{,}060,\ \$2{,}820{,}844]$, $N=1{,}000$ paired seeds) and mean breaches per iteration by $+0.158$ (95\% CI $[+0.084,\ +0.238]$). The analogous paired comparison against the non-binding band (budget$=5$ vs budget$=40$) yields $\Delta\mathrm{ALE}=+\$1{,}655{,}517$ (95\% CI $[\$495{,}551,\ \$2{,}940{,}651]$) and $\Delta\text{breaches}=+0.169$ (95\% CI $[+0.105,\ +0.239]$). Within the binding regime itself, budget$=10$ vs budget$=40$ is not statistically distinguishable ($\Delta\mathrm{ALE}=+\$172{,}730$, 95\% CI $[-\$987{,}521,\ +\$1{,}298{,}069]$) at $N=1{,}000$, indicating the transition is concentrated at the $\rho \approx 1$ edge rather than spread gradually across 5--20 hrs/mo.

\textit{(iii) Non-binding regime} ($\geq 40$ hr/mo, $\rho \ll 1$). Every variance event is fully remediated within its budget cycle and the queue empties before the next arrival. Additional budget hours go unspent and therefore cannot affect the trajectory --- which is why budgets of 40, 80, and 160 hours/month produce statistically equivalent results across seeds. The three regimes correspond to the classical utilization map of a single-server priority queue~\citep{Kleinrock1975}; this experiment localizes the transition at 5--10~hr/mo for this specific control portfolio.

\paragraph{Sensitivity to the remediation scheduling strategy.} All results above use the \code{budget\_efficiency} discipline (cheapest-first within priority tier). The simulator implements two additional disciplines (\texttt{perceived\_risk}, \texttt{worst\_case\_impact}); a preliminary cross-strategy sweep shows the scheduling discipline is a first-order risk factor whose effect exceeds the cross-budget effect at saturation, following the textbook SPT-vs-LPT pattern~\citep{Kleinrock1975}. A full cross-strategy study, including a crossed budget $\times$ scheduling design to assess interaction effects, is deferred to a dedicated follow-on paper.

\subsection{Scenario results}
\label{sec:scenario-results}

We ran both hospital ransomware scenarios (Section~\ref{sec:scenarios}) for 1{,}000 iterations at 1-year horizon (seeds 0--999) under default configuration (40 hrs/month remediation budget, budget\_efficiency scheduling).

\paragraph{Scenario A: Weak controls.} Nearly every iteration produces at least one breach (97.6\%), averaging 43.59 per year with a mean ALE of \$365.3M. Mean 94.8 contacts/year, 0 avoided, 51.2 resisted, 43{,}590 breaches across 1{,}000 iterations. Median ALE: \$204.4M. P95: \$1.30B. P99: \$1.93B. Maximum: \$5.96B. Mean variance events: 8.96. The scenario is designed to exercise the degraded-controls regime, not to estimate real-world incidence rates.

\paragraph{Scenario B: Medium controls.} Well-configured controls produce near-zero per-contact breach fractions: only 3.3\% of iterations produced any breach, with losses driven entirely by variance dynamics. Mean 54.6 contacts/year, 25.4 avoided, 29.1 resisted. 85 breaches across 1{,}000 iterations. Mean ALE: \$930{,}875. Mean breaches per iteration: 0.085. Median: \$0. P95: \$0 (96.7\% of iterations produce zero loss). P99: \$24.5M. Maximum: \$382.6M. Mean variance events: 12.66. With 96.7\% of iterations producing zero loss, the mean ALE is dominated by a small number of catastrophic iterations; the P99 is a more relevant metric than the mean or the P95 for board-level risk communication.

\paragraph{Scenario comparison.} The ratio of weak-to-medium mean ALE is approximately $392\times$ ($N=1{,}000$). The order-of-magnitude conclusion is robust but precise multiplicative claims are not, given the heavy-tailed nature of the medium scenario's loss distribution. The gap is driven by variance dynamics: the weak scenario's 46\% per-contact breach fraction far exceeds what its intended RS of $\sim$0.94 would predict, because slow VMC detection allows variance to accumulate undetected. The medium scenario's fast VMCs keep effective RS near its intended level.

Notably, the weak scenario has fewer \emph{detected} variance events (9.0 vs 12.7) because its slow VMCs rarely detect drift. More detected variance indicates healthier monitoring, a specific instance of Goodhart's Law~\citep{Goodhart1984} where the measurement regime dominates the metric.

\begin{figure}[htbp]
\centering
\includegraphics[width=0.95\textwidth]{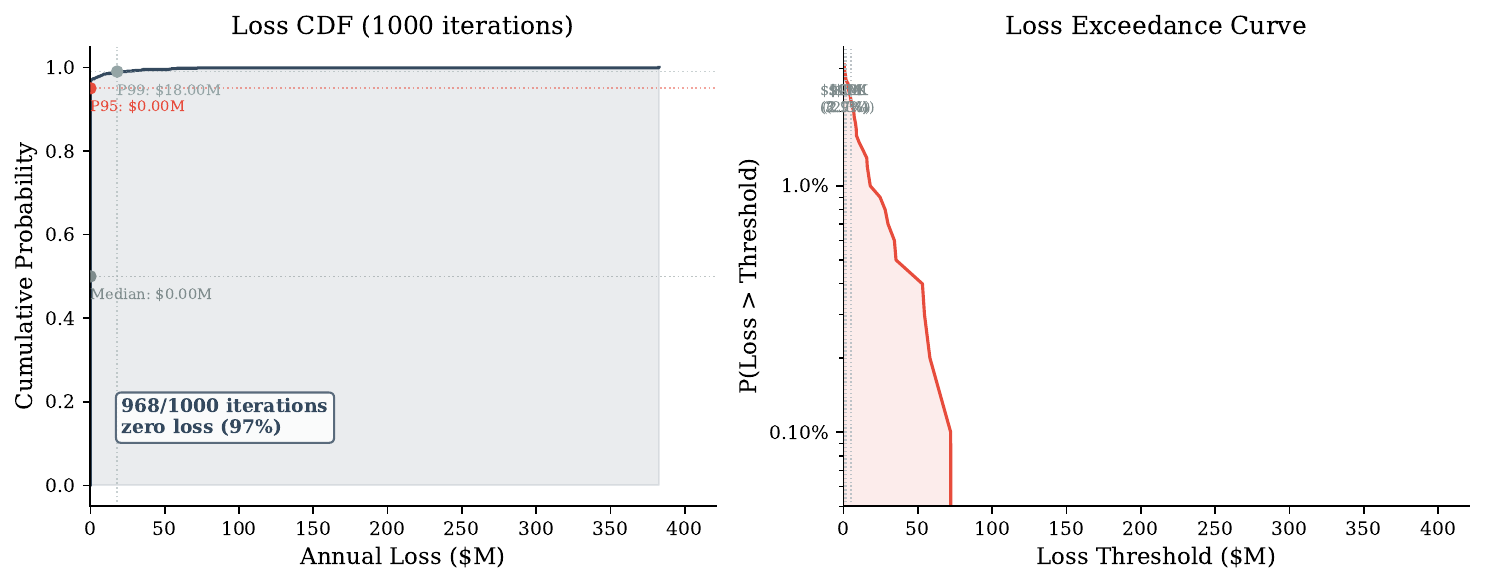}
\caption{Loss distribution for the medium controls scenario (1{,}000 iterations, 1-year horizon). Left: loss CDF showing 96.7\% of iterations produce zero loss (P95 $=\$0$, P99 $=\$24.5$M). Right: loss exceedance curve showing the extreme right skew characteristic of catastrophic risk.}
\label{fig:lossdist}
\end{figure}

\paragraph{Loss distribution characteristics.} The medium scenario's loss distribution is extremely heavy-tailed (Figure~\ref{fig:lossdist}): 96.7\% of iterations produce zero loss, but the worst produces \$382.6M --- 411$\times$ the mean ALE. The P99 is therefore a more relevant summary than mean or P95 for risk-appetite communication.

\begin{figure}[htbp]
\centering
\includegraphics[width=0.75\textwidth]{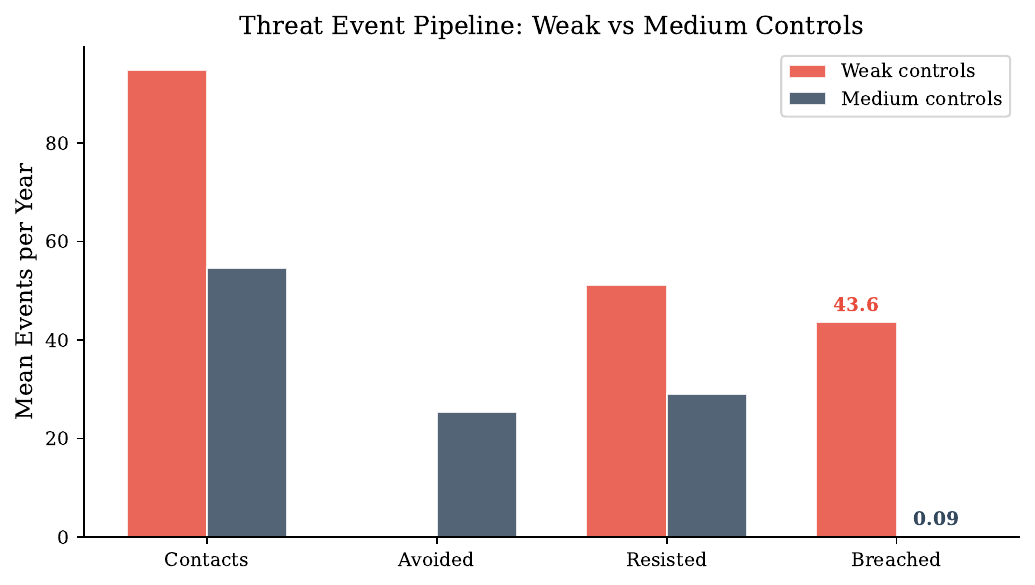}
\caption{Threat event pipeline comparison (1{,}000 iterations each, 1-year horizon). The weak scenario produces 43.59 breaches/year (46\% per-contact breach fraction) while the medium scenario produces 0.085 breaches/year (0.16\% per-contact breach fraction) --- an approximately $392\times$ difference in mean expected loss, bounding the value of organizational control quality across this scenario family. Note that the weak scenario shows a higher \emph{absolute} number of resisted threats (51.2 vs 29.1) because it has no avoidance controls to strip contacts at the prior gate, so more threats reach the resistance stage; its resistance \emph{rate} (51.2/94.8 $\approx$ 54\%) is nevertheless far below the medium scenario's ($\approx 99\%$).}
\label{fig:funnel}
\end{figure}

The threat event pipeline for both scenarios is shown in Figure~\ref{fig:funnel}. Scenario comparison is summarized in Table~\ref{tab:scenarios}.

\begin{table}[htbp]
\centering
\caption{Hospital ransomware scenario comparison (1{,}000 iterations, 1-year horizon, seeds 0--999). Bracketed values are 95\% nonparametric bootstrap confidence intervals (10{,}000 resamples).}
\label{tab:scenarios}
\small
\begin{tabular}{l r r}
\toprule
\textbf{Metric} & \textbf{Weak} & \textbf{Medium} \\
\midrule
Mean contacts/year      & 94.8  & 54.6 \\
Mean avoided            & 0.0   & 25.4 \\
Mean resisted           & 51.2  & 29.1 \\
Mean breaches/year      & 43.59 & 0.085 \\
Iterations with breaches & 976/1{,}000 & 33/1{,}000 \\
Mean ALE                & \$365{,}271{,}330 & \$930{,}875 \\
Median ALE              & \$204{,}413{,}751  & \$0 \\
P95 ALE                 & \$1{,}295{,}360{,}818  & \$0 \\
P99 ALE                 & \$1{,}927{,}129{,}192  & \$24{,}501{,}582 \\
Max ALE                 & \$5{,}955{,}716{,}283  & \$382{,}632{,}205 \\
Mean variance events    & 8.96   & 12.66 \\
Breach fraction         & 46.0\% & 0.16\% \\
\midrule
\multicolumn{3}{l}{\textit{Cross-scenario:} mean-ALE ratio Weak/Medium $= 392\times$.} \\
\bottomrule
\end{tabular}
\end{table}

\section{Limitations}
\label{sec:knownlim}

The following define the simulator's current boundaries.

\begin{enumerate}[nosep]
    \item \textbf{No comparison to simpler baselines.} We do not compare the ABM against simpler modeling approaches (e.g., a Markov chain of control states, or a discrete-event simulation without agent autonomy). Some findings (particularly the queueing regime transition) might be reproducible with a simpler model. The ABM's incremental value over such baselines lies in the narrative causation engine, the cascade dynamics, and the extensibility to personnel dynamics, but we have not quantified this value formally.

    \item \textbf{VMC variance detection is monolithic.} Each VMC samples a detection interval and applies a single efficacy when sweeping for variant controls. The full FAIR-CAM stage-gated decomposition into visibility, monitoring, and recognition sub-functions for VMC detection is not yet implemented; LEC breach detection (Equation~\ref{eq:detect-and}) does implement the AND-gate.

    \item \textbf{Coverage not explicitly modelled.} Coverage is determined by control wiring (effectively Cov = 1.0 for all linked controls). For scenarios requiring partial coverage (e.g., EDR deployed on 80\% of endpoints), this gap matters. Shadow IT is representable as a declared technology-asset class with no control wiring (Section~\ref{sec:abstraction}); implicit or probabilistic coverage is deferred to future work.

    \item \textbf{Incident-feedback loops and additional simplifications.} Breach events do not feed back into personnel awareness, budget allocation, or control prioritization; the budget-threshold effect in Section~\ref{sec:budgetregime} should be read as conditional on these loops being absent (see Section~\ref{sec:discussion}). Dynamic personnel behavior is disabled (static DSC attributes; closed-loop evaluation is a follow-on paper). A single organizational archetype and a single remediation scheduling strategy (\code{budget\_efficiency}) are exercised. Controls occupy discrete states (normal/variant/remediating) without partial degradation. The five DSC dimensions are modeled as independent Bernoulli trials, though in practice they co-vary through common causes.

    \item \textbf{Loss minimization percentage-based.} Loss reduction controls apply efficacy-based percentage reduction rather than the FAIR-CAM formula for insurance-type controls with caps and deductibles.

\end{enumerate}

\section{Discussion}
\label{sec:discussion}

\subsection{Implications for FAIR practitioners}

Static FAIR analysis with analytically computed OpEff is sufficient when controls are approximately independent, remediation budget comfortably exceeds demand, and VMC reliability is high. When any of these conditions fail, the ABM captures dynamics the formula cannot: cascading monitoring failures, non-linear budget thresholds, and correlated variance accumulation.

The narrative causation engine adds actionable root cause traceability. It transforms risk communication from ``our expected loss is \$2M'' to ``70\% of our losses are resource-constrained; increasing remediation budget from 10 to 40 hours/month would eliminate the backlog that enables them.'' We recommend a tiered approach: use analytical FAIR for initial risk screening across the portfolio, then apply ABM simulation to scenarios where control interdependencies, budget constraints, or organizational behavior are expected to produce non-linear effects. The analytical approach requires fewer inputs, runs instantly, and is easier to communicate; it is preferable when its stationarity assumptions hold.

\subsection{Connection to safety science}

The organizational dynamics observed in Section~\ref{sec:cascade} and Section~\ref{sec:budgetregime} instantiate patterns described in the safety science literature introduced in Section~\ref{sec:faircam}. The VMC cascade mechanism (Section~\ref{sec:cascade}) is a computational realization of Reason's latent conditions~\citep{Reason1990}: a VMC that degrades is a latent condition, undetected and persistent, that enables future active failures when threats arrive. The narrative causation engine (Section~\ref{sec:rca}) makes this latent-to-active transition observable: for each loss event it traces which VMCs were degraded, how long the degradation went undetected, and which controls accumulated exploitable variance during that window. The budget-driven variance accumulation (Section~\ref{sec:budgetregime}) mirrors Rasmussen's migration toward safety boundaries~\citep{Rasmussen1997}: under resource pressure the system drifts incrementally toward a regime where variance events compound faster than they are resolved, through locally rational prioritization decisions that are individually defensible. The simulation locates this boundary rather than merely asserting its existence. The monitoring topology's role in propagating correlated failure (a single VMC degradation silently affecting 70\% of the LEC population in the illustrative trace) is consistent with Perrow's observation that tightly coupled systems produce outcomes no single-component failure could explain~\citep{Perrow1984}. The contribution is the ability to parameterize, reproduce, and quantify these patterns within a single framework aligned to the FAIR-CAM standard.

\subsection{Future work}

Open directions include variance-based sensitivity analysis (Sobol indices) across portfolio size and organizational structure, systematic comparison of remediation scheduling disciplines, and closing the model's open causal chains into feedback loops. Whether these are pursued within the current architecture or through integration with complementary frameworks is an active design decision.

\section{Conclusion}

This paper presents the first computational implementation of the core FAIR-CAM dynamics and releases it for practitioners and researchers. In a hospital ransomware scenario, the implementation reveals three organizational dynamics that static analysis cannot represent: emergent operational efficacy diverges systematically from analytical values under correlated variance, the remediation pipeline exhibits a sharp queueing regime transition, and monitoring failures cascade through the VMC topology. The specific magnitudes are scenario-dependent; the general claim is the existence of a shared, inspectable, parameterizable tool against which both formulas and intuitions about organizational risk can now be checked.

\paragraph{Data and code availability.} The simulator source code, experimental scripts, and scenario configuration files are available at \url{https://github.com/security-decision-science/security-decision-labs/tree/main/tools/} (Python~3.10+, NumPy, SciPy, Mesa, NetworkX). All experiments use fixed RNG seeds for reproducibility; see \texttt{REPRODUCE.md} in the repository for script-to-figure mappings and exact reproduction commands.

\section*{Declaration of competing interest}
Jack Jones is the creator of the FAIR and FAIR-CAM frameworks and a co-founder of the FAIR Institute. Laura Voicu declares no competing financial interests.

% --- References ---
\bibliographystyle{elsarticle-num}

\end{document}